\documentclass[12pt,a4paper,superscriptaddress,preprint]{revtex4-1}
\usepackage[dvips]{graphicx}
\usepackage{amssymb,amsmath}

\newcommand{\FIGS}{Figs.~}
\newcommand{\SEC}{Sec.~}
\newcommand{\SECS}{Secs.~}

\begin{document}

\title{Voter models with contrarian agents}
\date{\today}
\author{Naoki Masuda}
\affiliation{Department of Mathematical Informatics,
The University of Tokyo,
7-3-1 Hongo, Bunkyo, Tokyo 113-8656, Japan}

\begin{abstract}
In the voter and many other opinion formation models, agents are assumed to behave as congregators (also called the conformists); they are attracted to the opinions of others. In this study, I investigate linear extensions of the voter model with contrarian agents. An agent is either congregator or contrarian and assumes a binary opinion. I investigate three models that differ in the behavior of the contrarian toward other agents. In model 1, contrarians mimic the opinions of other contrarians and oppose (i.e., try to select the opinion opposite to) those of congregators. In model 2, contrarians mimic the opinions of congregators and oppose those of other contrarians. In model 3, contrarians oppose anybody. In all models, congregators are assumed to like anybody. I show that even a small number of contrarians prohibits the consensus in the entire population to be reached in all three models. I also obtain the equilibrium distributions using the van Kampen small-fluctuation approximation and the Fokker-Planck equation for the case of many contrarians and a single contrarian, respectively. I show that the fluctuation around the symmetric coexistence equilibrium is much larger in model 2 than in models 1 and 3 when contrarians are rare.
\end{abstract}

\maketitle

\section{Introduction}\label{sec:introduction}

Dynamics of collective opinion formation is widely studied in various
disciplines including statistical physics. In typical
models of opinion formation,
agents interact and dynamically change the opinion, which I call the
state, according to others' states and perhaps the agent's own state.
The voter model is a paradigmatic stochastic model of this kind
\cite{Liggett1985book,Castellano2009RMP,Redner2001book,Krapivsky2010book}.
In the voter model, each agent flips the binary state at a rate
proportional to the number of neighboring agents possessing
the opposite state. In arbitrary finite contact networks and in 
some infinite networks, the stochastic dynamics of the voter model
always ends up with perfect consensus of either state. The time required
before the consensus is reached has been characterized in many cases.
The possibility of consensus and the relaxation time, among other things,
have also been examined in other opinion formation models \cite{Galam2008IJMPC,Castellano2009RMP}. 

The voter model as well as many other opinion formation models
assume that the population is homogeneous. In
fact, real agents are considered to be heterogeneous in various
aspects. The agents' heterogeneity has been incorporated into the
voter model in the form of, for example, heterogeneous degrees (i.e., number of
neighbors) in the contact network
\cite{Sood2005PRL,Suchecki2005PRE,Suchecki2005EPL,Sood2008PRE,Antal2006PRL,MasudaOhtsuki2009NewJPhys}, positions in the
so-called Watts-Strogatz small-world network
\cite{Castellano2003EPL,Vilone2004PRE,Suchecki2005PRE},
heterogeneity in the flip rate
\cite{MasudaGibertRedner2010PRE,MasudaRedner2011JSM}, 
and zealosity \cite{Galam1991EurJSocPsychol,Mobilia2003PRL,Mobilia2007JSM}.

In the present study, I examine extensions of the voter model in
which some agents are not like-minded voters. Such contrarian agents would
transit to the state opposite to that of others and were first studied in Ref.~\cite{Galam2004PhysicaA}.
It should be noted that
contrarians are assumed to dynamically change their states; contrarians are assumed to be zealots (i.e., those that never change the state) in a previous study \cite{LiBraunstein2011PRE}.
In models in which consensus is the norm 
in the absence of
contrarians, contrarians often prohibit the consensus to be
reached such that the dynamics finally reaches the coexistence of
different states. This holds true for
the majority vote model
\cite{Galam2004PhysicaA,Stauffer2004PhysicaA,Borghesi2006PRE,Galam2007QualQuant,LiBraunstein2011PRE}, Ising
model \cite{Kurten2008IJMPB}, so-called Sznajd model
\cite{delaLama2005EPL}, a model with a continuous state space
\cite{Martins2010ACS}, and a general model including some of these
models \cite{Nyczka2013JSP}.
These models show phase transitions
between a consensus (or similar) phase and a coexistence
phase when the fraction of contrarians in the population (i.e.,
quenched randomness) \cite{Stauffer2004PhysicaA,Martins2010ACS,Kurten2008IJMPB}
or the probability of the contrarian behavior adopted by all the
agents in the population
(i.e., annealed randomness) is varied \cite{Galam2004PhysicaA,Stauffer2004PhysicaA,Borghesi2006PRE,delaLama2005EPL,Nyczka2013JSP}. The effects of contrarians have been also examined in the so-called minority game \cite{Zhong2005PRE_minority}.

In contrast to these nonlinear models, I focus on three linear extensions
of the voter model with contrarian agents (i.e., quenched
randomness). By linearity, I mean to pertain to stochastic mass interaction.
A previous study numerically examined coevolutionary dynamics of a
linear extension of the voter model with contrarians and network formation
\cite{YiBaekZhuKim2013PRE}. In contrast, 
I focus on a fixed and well-mixed population.
I show that even a small density of contrarians changes the
collective dynamics of the extended voter models from the consensus
configuration to the coexistence configuration.
I also analytically quantify the fluctuations in the agents' behavior
in the coexistence equilibrium.

\section{Model}

I consider three variants of the voter model with contrarians. 
The agent that obeys the state transition rule of the
standard voter in the voter model is referred to as
congregator.
The fraction of congregators and that of contrarians are denoted by
$X$ and $Y (=1-X)$, respectively.
Contrarian is assumed to be a quenched property. In other words, an agent is either congregator or contrarian throughout the dynamics.
Each agent, either congregator or contrarian, takes either
state ${\mathbf 0}$ or state ${\mathbf 1}$ at any time.
I denote the mean fraction of congregators
in state ${\mathbf 1}$ within the congregator subpopulation by $x$
and that of contrarians
in state ${\mathbf 1}$ within the contrarian subpopulation by $y$
($0\le x, y\le 1$).
The mean fractions of congregators and contrarians in state ${\mathbf 0}$ in the congregator and contrarian subpopulations are given by $1-x$ and $1-y$, respectively.

I assume that the population is well mixed and contains $N$ agents.
The continuous-time stochastic opinion dynamics is defined as follows. Each congregator in state ${\mathbf 0}$ independently flips to state ${\mathbf 1}$ with the rate equal to the number of ${\mathbf 1}$ agents, no matter whether they are congregators and contrarians. Likewise, each congregator in state ${\mathbf 1}$ flips to state ${\mathbf 0}$ with the rate equal to the number of ${\mathbf 0}$ agents. This assumption is common to the three models.
The behavior of the congregator in the present models is the same as that of
the voter in the standard voter model.

The three models are different in the behavior of contrarians as follows.
In model 1, it is assumed that contrarians oppose
congregators and like contrarians. In other words, each contrarian in state ${\mathbf 0}$
independently flips to state ${\mathbf 1}$ with the rate equal to the
sum of the number of ${\mathbf 0}$ congregators and
that of ${\mathbf 1}$ contrarians.
In model 2, contrarians like congregators and oppose contrarians.
In other words, each contrarian in state ${\mathbf 0}$ flips to state
${\mathbf 1}$ with the rate equal to the sum of the number of ${\mathbf 1}$ congregators and that of ${\mathbf 0}$ contrarians.
In model 3, contrarians oppose both congregators and contrarians.
In other words, each contrarian in state ${\mathbf 0}$ flips to state
${\mathbf 1}$ with the rate equal to the sum of the number of ${\mathbf 0}$ agents.
In all models, the parallel definition is applied to the flip rate for
the contrarian to transit from ${\mathbf 1}$ to ${\mathbf 0}$. 
It should be noted that the cognitive demand for the agents is considered to be the lowest for model 3 because the contrarian does not have to recognize
the type of other agents when possibly updating its state.
The
definition of the three models is summarized in Table~\ref{tab:3 models def}.

\begin{table}[h]
\caption{Agents' behavior in the three models.}
\label{tab:3 models def}
\begin{center}
\begin{tabular}{|c|c|c|c|}\hline
Behavior & Model 1 & Model 2 & Model 3\\ \hline
Congregators toward congregators & Like & Like & Like\\
Congregators toward contrarians & Like & Like & Like\\
Contrarians toward congregators & Oppose & Like & Oppose\\
Contrarians toward contrarians & Like & Oppose & Oppose\\ \hline
\end{tabular}
\end{center}
\end{table}

\section{Results}

\subsection{Mean-field dynamics}\label{sub:rate equations}

The rate equations for model 1 are given by
\begin{align}
\frac{{\rm d}x}{{\rm d}t} =& (1-x)(Xx+Yy) - x\left[X(1-x)+Y(1-y)\right],\label{eq:x}\\
\frac{{\rm d}y}{{\rm d}t} =& (1-y)\left[X(1-x)+Yy\right]
- y\left[Xx+Y(1-y)\right]. \label{eq:y model 1}
\end{align}
If $0<Y<1$, the steady state is given by
\begin{equation}
(x^*, y^*) = \left(\frac{1}{2}, \frac{1}{2}\right),
\label{eq:stationary}
\end{equation}
where $*$ denotes the values in the equilibrium.
It should be noted that putting $X=1$ and $Y=1-X=0$ in Eq.~\eqref{eq:x} yields the standard voter model. In this case, we obtain ${\rm d}x/{\rm d}t=0$, which implies that $x$ is conserved. It is an artefact of the mean-field equation. In fact, stochastic dynamics of the voter model drives the population toward $x=0$ or $x=1$, which are absorbing configurations. In contrast, $x=0$ and $x=1$ with whatever $y$ values are not absorbing in the present model with $0<Y<1$.

Using the relationship $X+Y=1$, the following
characteristic equation is obtained for the mean-field dynamics in the steady state given by Eq.~\eqref{eq:stationary}:
\begin{equation}
\lambda^2 + \lambda + 2 Y(1-Y) = 0
\label{eq:eigen eq model 1}
\end{equation}
Because the real parts of the two eigenvalues obtained from Eq.~\eqref{eq:eigen eq model 1}
are negative, the steady state is stable.

Therefore, consensus is not asymptotically reached in this model, and the dynamics starting from an arbitrary initial condition tends to the steady state given by
Eq.~\eqref{eq:stationary}, regardless of the density of contrarians, $Y$.
If $Y\ge (2+\sqrt{2})/4$ or $Y\le (2-\sqrt{2})/4$,
the two eigenvalues are
real, such that the dynamics overdamps to the equilibrium.
If $(2-\sqrt{2})/4<Y<(2+\sqrt{2})/4$,
the two eigenvalues have imaginary parts such that the relaxation accompanies an oscillation.

The equilibrium fraction of agents in either state is equal to $1/2$, for both the congregator subpopulation and contrarian subpopulation, regardless of the density of contrarians in the population. The influence of even just a few number of contrarians on the dynamics can be huge; they prevent the consensus. The behavior of the model is very different from that of the 
voter model, for which consensus is necessarily reached via diffusion.

In the limit $Y\ll 1$, the larger eigenvalue, which
determines the decay rate of the dynamics to the steady state,
is approximately equal to $\approx -2Y$. Therefore, for a small density of contrarian, the actual dynamics would fluctuate around the steady state in a long run. I will quantify fluctuations of the stochastic dynamics in \SECS\ref{sub:van Kampen} and \ref{sub:1 contrarian}.

For model 2, the rate equations are given by Eq.~\eqref{eq:x} and
\begin{equation}
\frac{{\rm d}y}{{\rm d}t} = (1-y)\left[Xx+Y(1-y)\right]
 - y\left[X(1-x)+Yy\right]. 
\label{eq:y model 2}
\end{equation}
The equilibrium is in fact given by Eq.~\eqref{eq:stationary}.
The characteristic equation in the equilibrium is 
given by
\begin{equation}
\lambda^2 + (1+2Y)\lambda + 2Y^2=0,
\label{eq:eigen eq model 2}
\end{equation}
which has two eigenvalues with negative real parts, implying that the equilibrium given by Eq.~\eqref{eq:stationary} is stable.
However, the leading eigenvalue when $Y\ll 1$ is given by $\lambda\approx -2Y^2$, which is much closer to zero than for model 1 (i.e., $\lambda\approx -2Y$). Therefore, the fluctuation in the equilibrium for model 2 is expected to be much larger than that for model 1. This is in fact the case, as shown in \SEC\ref{sub:van Kampen}.

For model 3, the rate equations are given by Eq.~\eqref{eq:x} and
\begin{equation}
\frac{{\rm d}y}{{\rm d}t} = (1-y)\left[X(1-x)+Y(1-y)\right] - y(Xx+Yy). 
\label{eq:y model 3}
\end{equation}
The equilibrium is again given by Eq.~\eqref{eq:stationary}.
The characteristic equation in the equilibrium is 
given by
\begin{equation}
\lambda^2 + (1+2Y)\lambda + 2Y=0,
\label{eq:eigen eq model 3}
\end{equation}
which has two eigenvalues with negative real parts.
Therefore, the equilibrium given by Eq.~\eqref{eq:stationary} is stable.
When $Y\ll 1$, the eigenvalue scales as $\lambda\approx -2Y$, the same as for model 1.

\subsection{van Kampen small-fluctuation approximation}\label{sub:van Kampen}

To understand the fluctuation around the equilibrium of the mean-field dynamics,
I carry out the small-fluctuation approximation of the master equation developed by van Kampen
\cite{Krapivsky2010book,Vankampenbook} for the three models. The van Kampen expansion reveals the relationship between the system size $N$ and the magnitude of fluctuation under the Gaussian assumption of the quantities of interest.
%
 
To this end, let us shift from the density description used in \SEC\ref{sub:rate equations}
to the number description. The number of congregators and that of contrarians
are denoted by $N_x$ and $N_y$, respectively. Let $n_x$ and $n_y$ represent the number of state ${\mathbf 1}$ congregators and that of state ${\mathbf 1}$ contrarians, respectively. It should be noted that $N=N_x+N_y$,
$0\le n_x\le N_x$, and $0\le n_y\le N_y$.
The ansatz for the van Kampen small-fluctuation approximation is given by
\begin{align}
n_x(t) =& N_x x(t) + \sqrt{N_x}\xi,
\label{eq:n_x ansatz}\\
n_y(t) =& N_y y(t) + \sqrt{N_y}\eta,
\label{eq:n_y ansatz}
\end{align}
where $x$ and $y$ are the mean densities of state ${\mathbf 1}$ congregators and state ${\mathbf 1}$ contrarians in the congregator and contrarian subpopulations, respectively, as introduced in \SEC\ref{sub:rate equations}.
$\xi$ and $\eta$ are stochastic variables, which are assumed to be intensive quantities.
I represent the probability that there are $n_x$ state ${\mathbf 1}$ congregators and $n_y$ state ${\mathbf 1}$ contrarians by
$P(n_x, n_y, t)=\Pi(\xi,\eta,t)$. 

\subsubsection{Model 1}

For model 1, the master equation in terms of $P$ is given by
\begin{align}
N\frac{{\rm d}P}{{\rm d}t} =&
(E_x-1)
\left[n_x\left(N_x-n_x+N_y-n_y\right)P\right]
+(E_x^{-1}-1)\left[\left(N_x-n_x\right)\left(n_x+n_y\right)P\right]\notag\\
+&(E_y-1)\left[n_y\left(n_x+N_y-n_y\right)P\right]
+(E_y^{-1}-1)\left[\left(N_y-n_y\right)\left(N_x-n_x+n_y\right)P\right],
\label{eq:master equation with P model 1}
\end{align}
where $E_x$, $E_x^{-1}$, $E_y$, and
$E_y^{-1}$ are the operators representing an increment in
$N_x$ by one, a decrement in $N_x$ by one, an increment in
$N_y$ by one, and a decrement in $N_y$ by one,
respectively. For example, the first term on the right-hand side of
Eq.~\eqref{eq:master equation with P model 1} represents the inflow
and outflow of the probability induced by a decrement in $N_x$
by one. The operators are given by
\begin{align}
E_x =& 1 + \frac{1}{\sqrt{N_x}}\frac{\partial}{\partial\xi}
+ \frac{1}{2N_x}\frac{\partial^2}{\partial\xi^2}+\cdots,
\label{eq:E_x}\\
E_x^{-1} =& 1 - \frac{1}{\sqrt{N_x}}\frac{\partial}{\partial\xi}
+ \frac{1}{2N_x}\frac{\partial^2}{\partial\xi^2}+\cdots,
\label{eq:E_x^-}\\
E_y =& 1 + \frac{1}{\sqrt{N_y}}\frac{\partial}{\partial\eta}
+ \frac{1}{2N_y}\frac{\partial^2}{\partial\eta^2}+\cdots,
\label{eq:E_y}\\
E_y^{-1} =& 1 - \frac{1}{\sqrt{N_y}}\frac{\partial}{\partial\eta}
+ \frac{1}{2N_y}\frac{\partial^2}{\partial\eta^2}+\cdots.
\label{eq:E_y^-}
\end{align}

By substituting Eqs.~\eqref{eq:n_x ansatz}, \eqref{eq:n_y ansatz},
\eqref{eq:E_x}, \eqref{eq:E_x^-}, \eqref{eq:E_y}, and \eqref{eq:E_y^-}
in Eq.~\eqref{eq:master equation with P model 1} and replacing the time derivative of $P$ by that of $\Pi$, I obtain
\begin{align}
& N\left(\frac{\partial\Pi}{\partial t} - \sqrt{N_x}\frac{{\rm d}x}{{\rm d}t}
\frac{\partial\Pi}{\partial\xi} - \sqrt{N_y}\frac{{\rm d}y}{{\rm d}t}
\frac{\partial\Pi}{\partial\eta} \right)\notag\\
=& \left(\frac{1}{\sqrt{N_x}}\frac{\partial}{\partial\xi}
+\frac{1}{2N_x}\frac{\partial^2}{\partial\xi^2} \right)
\left(N_x x+\sqrt{N_x}\xi\right)
\left[N_x\left(1-x\right)-\sqrt{N_x}\xi
+N_{y}\left(1-y\right)-\sqrt{N_y}\eta\right]\Pi\notag\\
&+
 \left(-\frac{1}{\sqrt{N_x}}\frac{\partial}{\partial\xi}
+\frac{1}{2N_x}\frac{\partial^2}{\partial\xi^2} \right)
\left[N_x\left(1-x\right)-\sqrt{N_x}\xi\right]
\left(N_x x+\sqrt{N_x}\xi
+N_{y}y+\sqrt{N_y}\eta\right)\Pi\notag\\
&+ \left(\frac{1}{\sqrt{N_y}}\frac{\partial}{\partial\eta}
+\frac{1}{2N_y}\frac{\partial^2}{\partial\eta^2} \right)
\left(N_yy+\sqrt{N_y}\eta\right)
\left[N_xx+\sqrt{N_x}\xi
+N_{y}\left(1-y\right)-\sqrt{N_y}\eta\right]\Pi\notag\\
&+ \left(-\frac{1}{\sqrt{N_y}}\frac{\partial}{\partial\eta}
+\frac{1}{2N_y}\frac{\partial^2}{\partial\eta^2} \right)
\left[N_y\left(1-y\right)-\sqrt{N_y}\eta\right]
\left(N_x\left(1-x\right)-\sqrt{N_x}\xi
+N_yy+\sqrt{N_y}\eta\right)\Pi. 
\label{eq:master equation with Pi model 1}
\end{align}

The highest order terms on the right-hand side, where $N_x$ and $N_y$ are regarded to be of the order of $N$, are equal to
\begin{align}
\sqrt{N_x}N_y
\left[x\left(1-y\right)-\left(1-x\right)y\right]\frac{\partial\Pi}{\partial\xi}
+
N_x\sqrt{N_y}
\left[xy-\left(1-x\right)\left(1-y\right)\right]\frac{\partial\Pi}{\partial\eta}.
\label{eq:highest model 1}
\end{align}
By comparing Eq.~\eqref{eq:highest model 1} to the highest order terms on the left-hand side of
Eq.~\eqref{eq:master equation with Pi model 1}, I obtain
\begin{align}
\frac{{\rm d}x}{{\rm d}t} =& \frac{N_y}{N}
\left[\left(1-x\right)y - x\left(1-y\right)\right],\label{eq:MF rho_x}\\
\frac{{\rm d}y}{{\rm d}t} =& \frac{N_x}{N}
\left[\left(1-x\right)\left(1-y\right) - xy\right],\label{eq:MF rho_y}
\end{align}
which are equivalent to
the mean-field dynamics given by Eqs.~\eqref{eq:x} and \eqref{eq:y model 1}.

By equating the second highest order terms in Eq.~\eqref{eq:master equation with Pi model 1}, I obtain
\begin{align}
N\frac{\partial\Pi}{\partial t} =&
N_y\frac{\partial}{\partial\xi}(\xi\Pi)
-\sqrt{N_xN_y}\eta\frac{\partial\Pi}{\partial\xi}
+\left[N_xx(1-x)+\frac{N_y}{2}(x+y-2xy)\right]
\frac{\partial^2\Pi}{\partial\xi^2}\notag\\
+& N_x\frac{\partial}{\partial\eta}(\eta\Pi) + \sqrt{N_xN_y}\xi\frac{\partial\Pi}{\partial\eta}
+\left[\frac{N_x}{2}(1-x-y+2xy)+
N_yy(1-y)\right]\frac{\partial^2\Pi}{\partial\eta^2}.
\label{eq:2nd highest model 1}
\end{align}

Application of $\int\int {\rm d}\xi {\rm d}\eta\,\xi$ and
$\int\int {\rm d}\xi {\rm d}\eta\,\eta$ to
Eq.~\eqref{eq:2nd highest model 1} yields
\begin{align}
N\frac{\partial\left<\xi\right>}{\partial t} =& -N_y\left<\xi\right> +
\sqrt{N_xN_y}\left<\eta\right>
\label{eq:d<xi>/dt model 1}
\end{align}
and
\begin{align}
N\frac{\partial\left<\eta\right>}{\partial t} =&
- \sqrt{N_xN_y}\left<\xi\right> - N_x\left<\eta\right>,
\label{eq:d<eta>/dt model 1}
\end{align}
respectively.
Because the characteristic equation for the
Jacobian of the dynamics given by Eqs.~\eqref{eq:d<xi>/dt
  model 1} and \eqref{eq:d<eta>/dt model 1} coincides with
Eq.~\eqref{eq:eigen eq model 1},
$\left<\xi\right>$ and
$\left<\eta\right>$ converge to the unique equilibrium given by
$\left<\xi\right>^* = \left<\eta\right>^*=0$.

Application of $\int\int{\rm d}\xi {\rm d}\eta\,\xi^2$,
$\int\int{\rm d}\xi {\rm d}\eta\,\xi\eta$, and
$\int\int {\rm d}\xi {\rm d}\eta\,\eta^2$
to Eq.~\eqref{eq:2nd highest model 1} yields
\begin{align}
N\frac{\partial\left<\xi^2\right>}{\partial t} =&
2N_xx(1-x) + N_y(x+y-2xy)
-2N_y\left<\xi^2\right> + 2\sqrt{N_xN_y}\left<\xi\eta\right>,
\label{eq:d<xi^2>/dt model 1}\\
N\frac{\partial\left<\xi\eta\right>}{\partial t} =&
-\sqrt{N_xN_y}\left<\xi^2\right>
-N\left<\xi\eta\right>+\sqrt{N_xN_y}\left<\eta^2\right>,
\label{eq:d<xi*eta>/dt model 1}
\end{align}
and
\begin{align}
N\frac{\partial\left<\eta^2\right>}{\partial t} =&
N_x(1-x-y+2xy) + 2N_yy(1-y)
- 2\sqrt{N_xN_y}\left<\xi\eta\right> -2N_x\left<\eta^2\right>,
\label{eq:d<eta^2>/dt model 1}
\end{align}
respectively.
By substituting $(x^*, y^*) = (1/2, 1/2)$ in
Eqs.~\eqref{eq:d<xi^2>/dt model 1}, \eqref{eq:d<xi*eta>/dt model 1},
and \eqref{eq:d<eta^2>/dt model 1} and setting the left-hand sides to 0, I obtain
\begin{align}
\left<\xi^2\right>^* =& \frac{N_x+3N_y}{8N_y},
\label{eq:<xi^2>^* model 1}\\
\left<\xi\eta\right>^* =& \frac{1}{8}\left(\sqrt{\frac{N_y}{N_x}}
-\sqrt{\frac{N_x}{N_y}}\right),
\label{eq:<xi*eta>^* model 1}\\
\left<\eta^2\right>^* =& \frac{3N_x+N_y}{8N_x}.
\label{eq:<eta^2>^* model 1}
\end{align}

In terms of the original variables, I obtain
\begin{align}
n_x =& \frac{N_x}{2} + \sqrt{N_x}\xi
\end{align}
and
\begin{align}
n_y =& \frac{N_y}{2} + \sqrt{N_y}\eta
\end{align}
in the infinite time limit.
Therefore, in terms of the fraction of ${\mathbf 1}$ congregators in
the congregator subpopulation and that of ${\mathbf 1}$ contrarians in the contrarian subpopulation, I obtain
\begin{align}
\sigma(x) = \frac{\sigma(n_x)}{N_x}
=\sqrt{\frac{\left<\xi^2\right>}{N_x}}
=\sqrt{\frac{N_x+3N_y}{8N_xN_y}}\label{eq:std x1},\\
\sigma(y) = \frac{\sigma(n_y)}{N_y}
=\sqrt{\frac{\left<\eta^2\right>}{N_y}}
=\sqrt{\frac{3N_x+N_y}{8N_xN_y}},\label{eq:std y1}
\end{align}
where $\sigma$ stands for the standard deviation.

\begin{figure}
\begin{center}
\includegraphics[width=8cm]{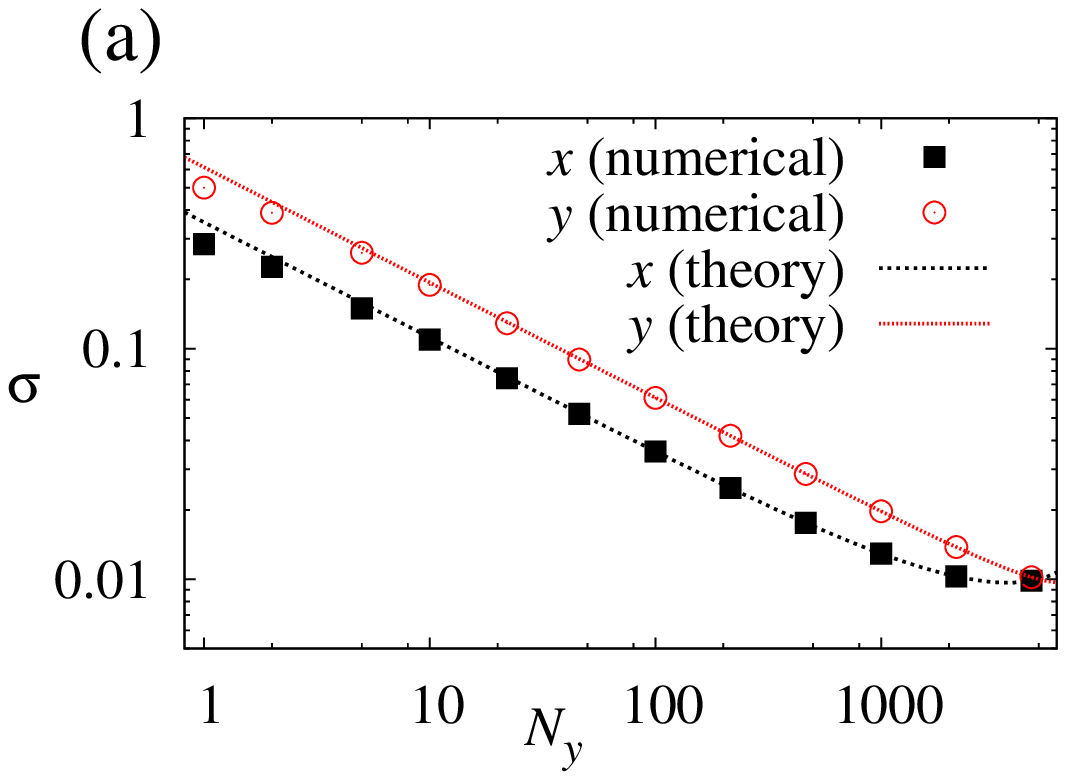}
\includegraphics[width=8cm]{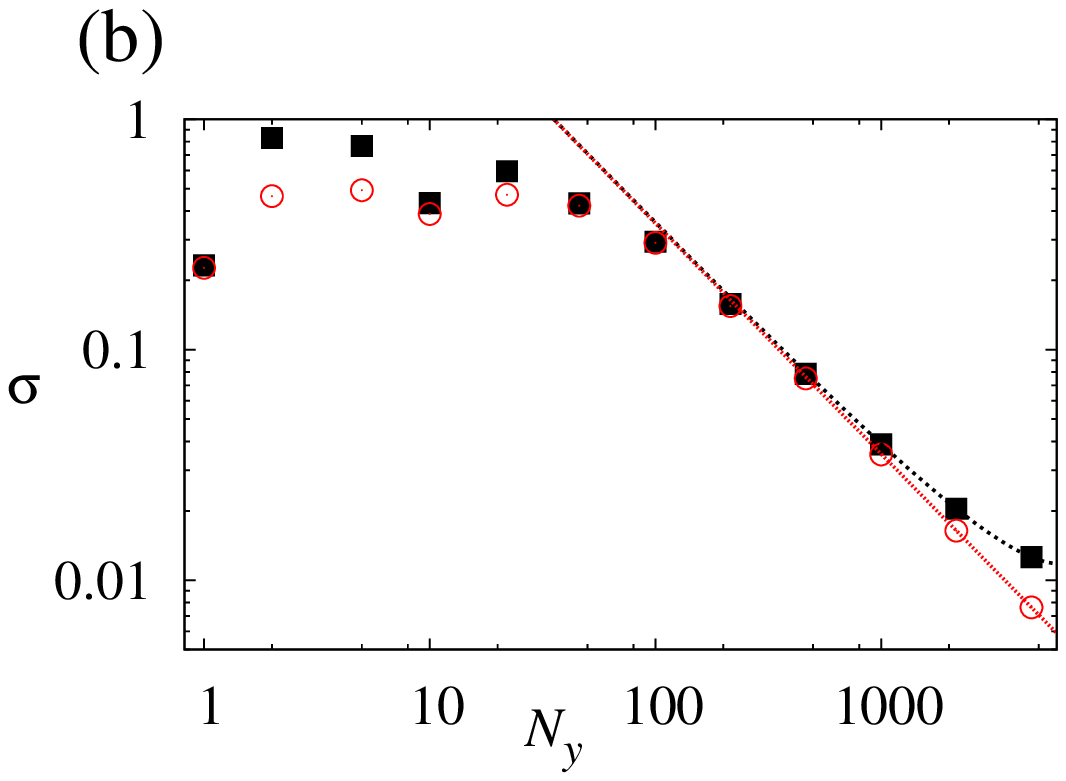}
\includegraphics[width=8cm]{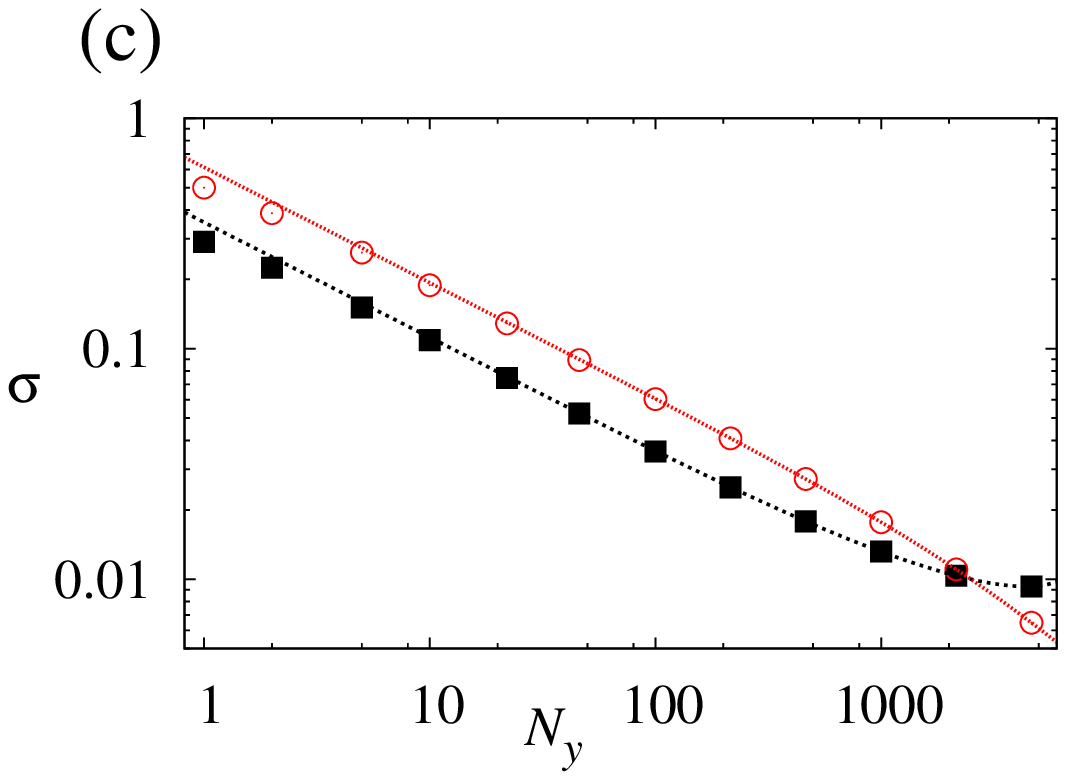}
\caption{The standard deviation in the fraction of congregators 
in state
  ${\mathbf 1}$ within the congregator subpopulation (i.e., $x$) and
that in the fraction of contrarians in state ${\mathbf 1}$ within the contrarian subpopulation (i.e., $y$). I set $N=10000$ and varied $N_y$.
The distributions are calculated on the basis of the results from
$t=0.5\times 10^7$ through $t=10^7$ in a single run starting
from $x=y=0.5$. This condition is common to the following numerical
results unless otherwise stated.
(a) Model 1; (b) model 2; (c) model 3.}
\label{fig:fluctuation}
\end{center}
\end{figure}

\begin{figure}
\begin{center}
\includegraphics[width=5cm]{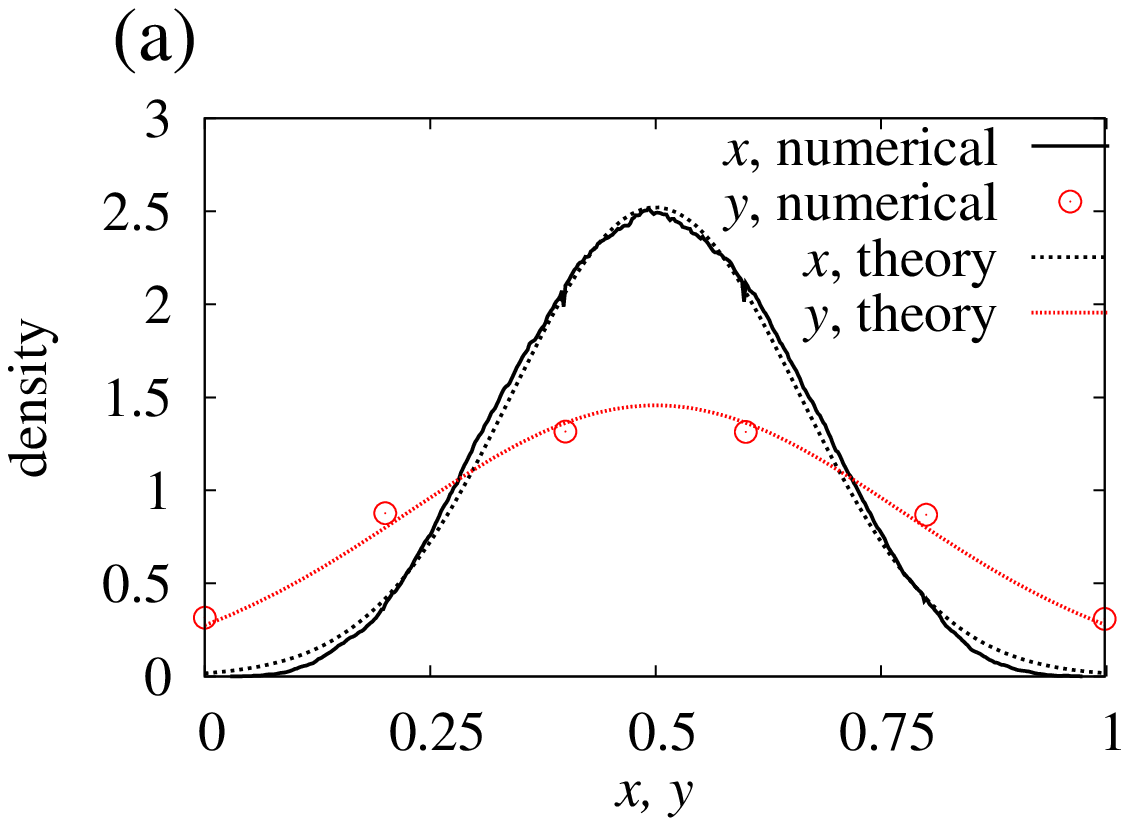}
\includegraphics[width=5cm]{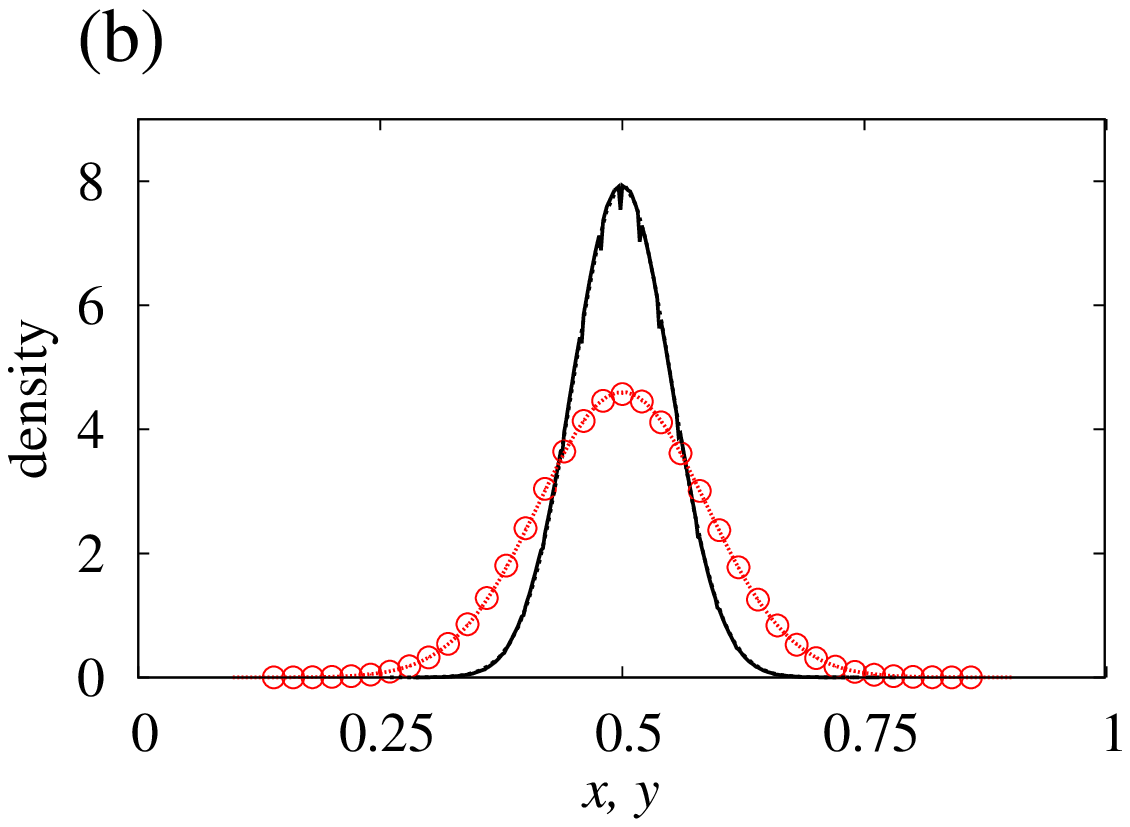}
\includegraphics[width=5cm]{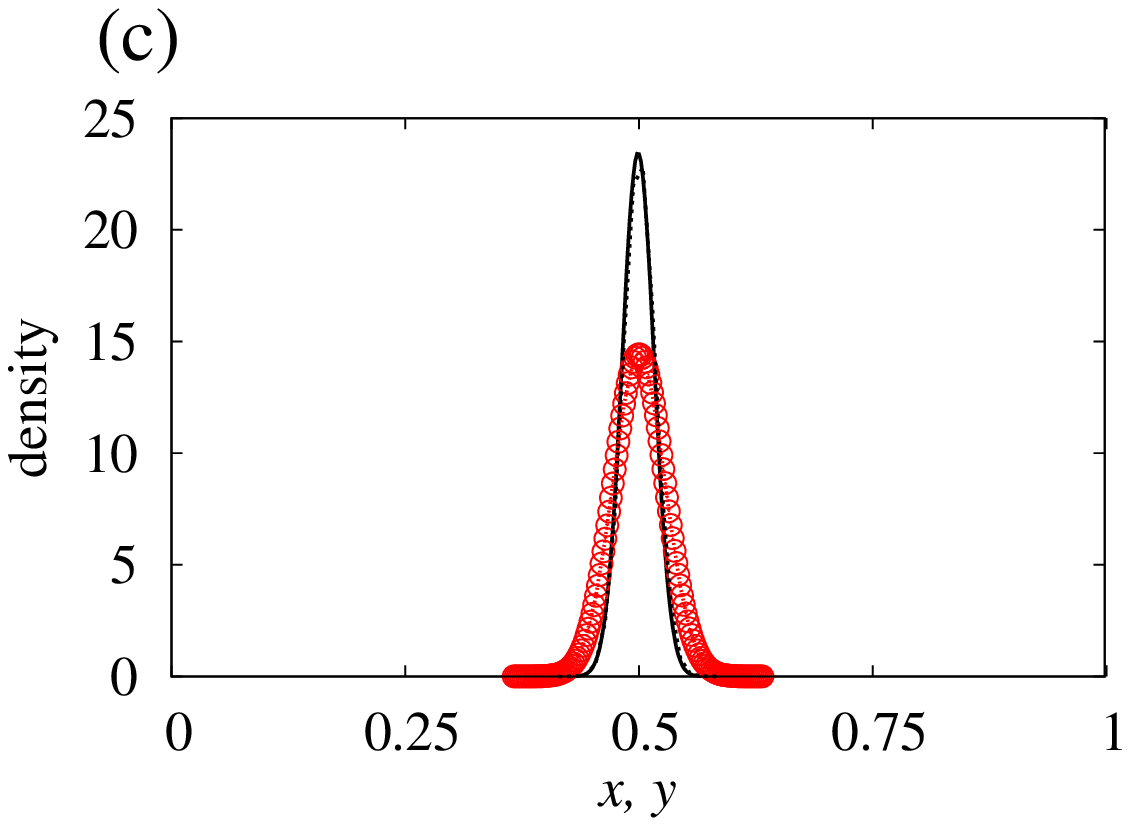}
\includegraphics[width=5cm]{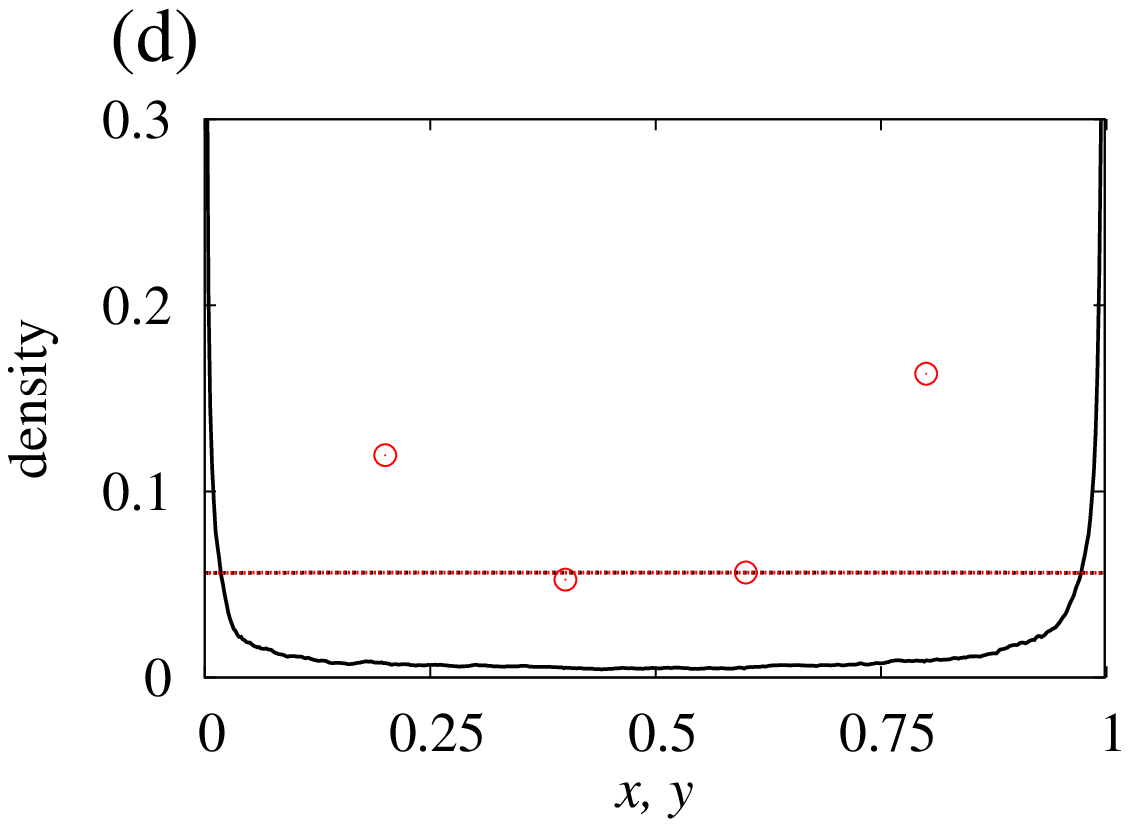}
\includegraphics[width=5cm]{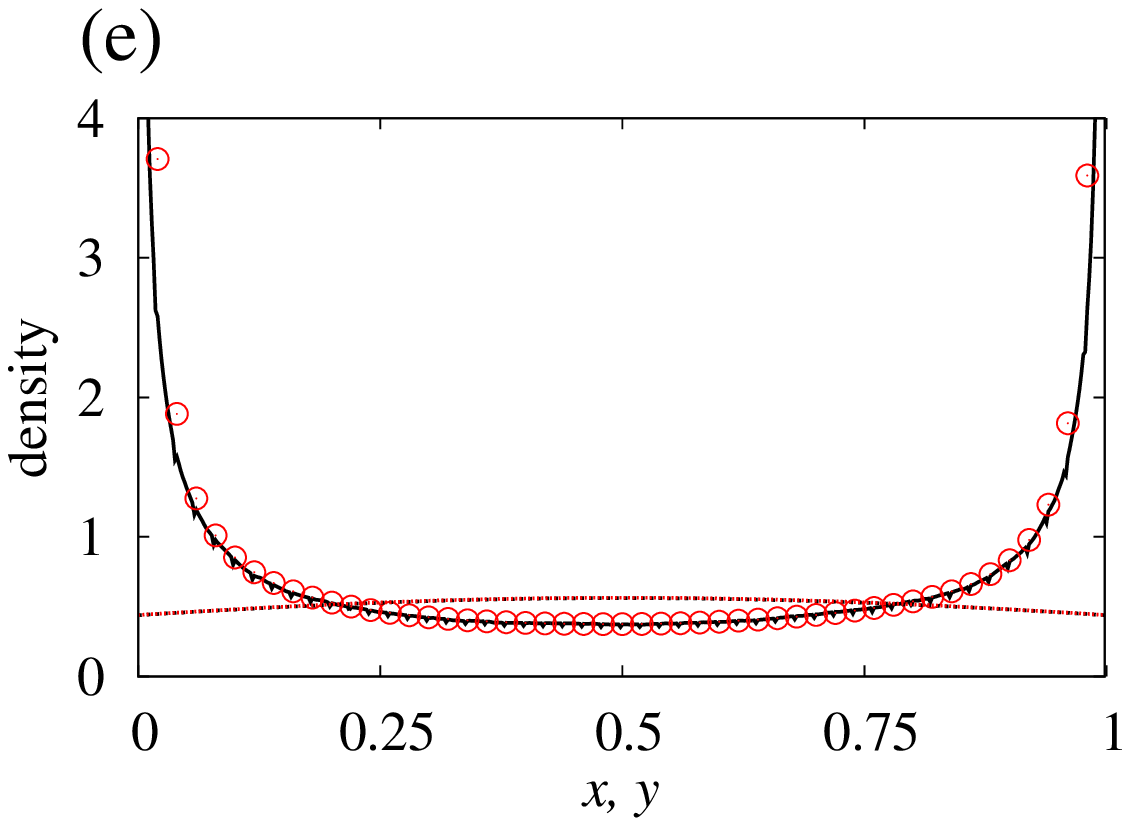}
\includegraphics[width=5cm]{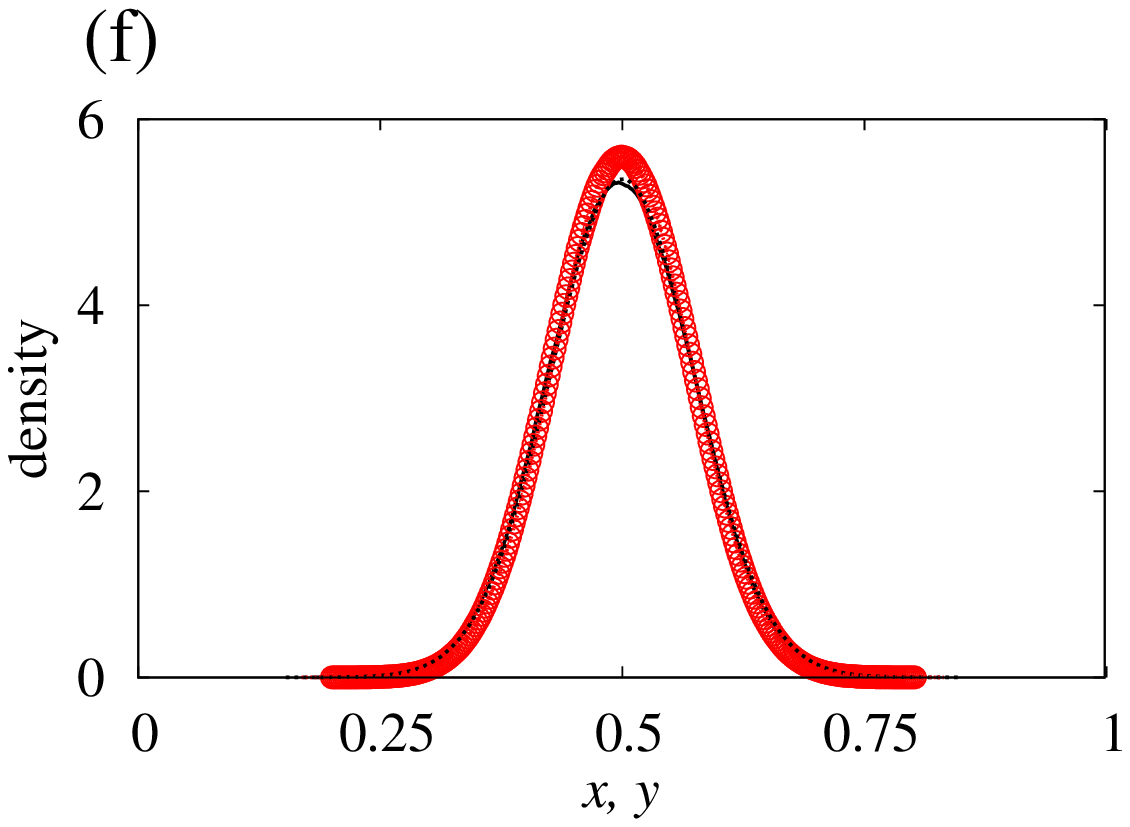}
\includegraphics[width=5cm]{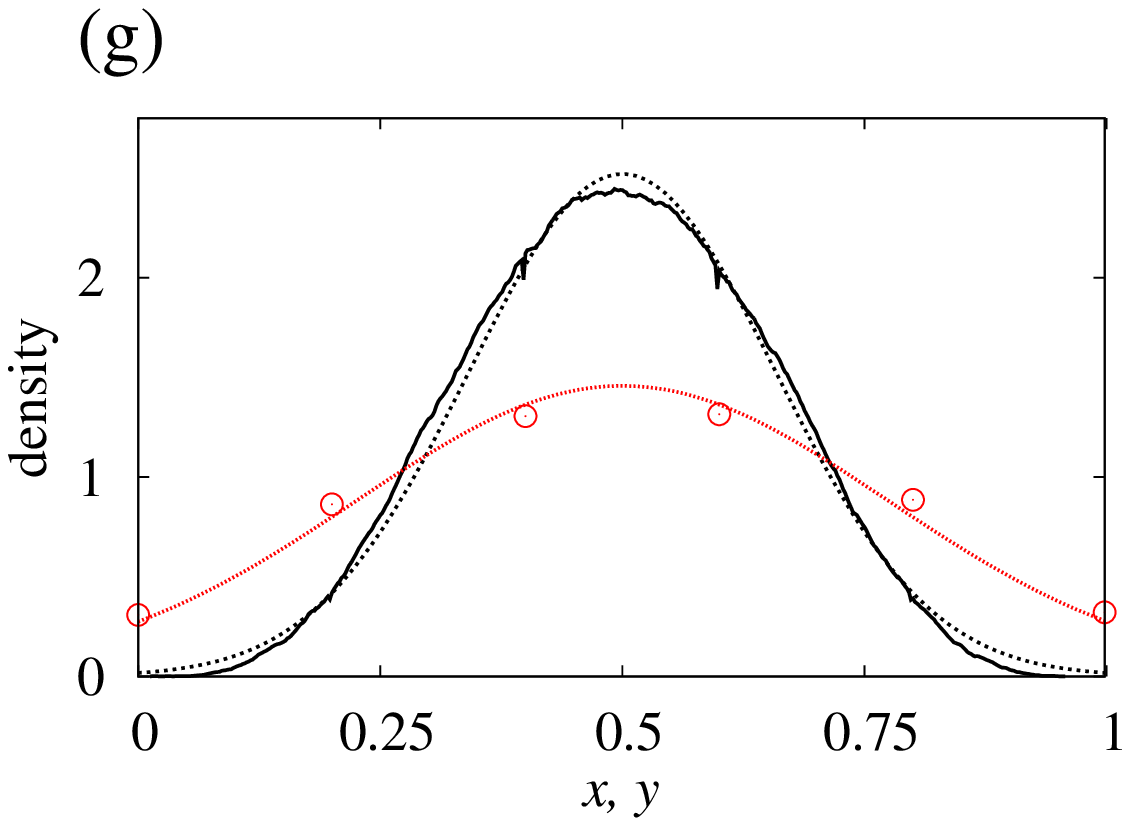}
\includegraphics[width=5cm]{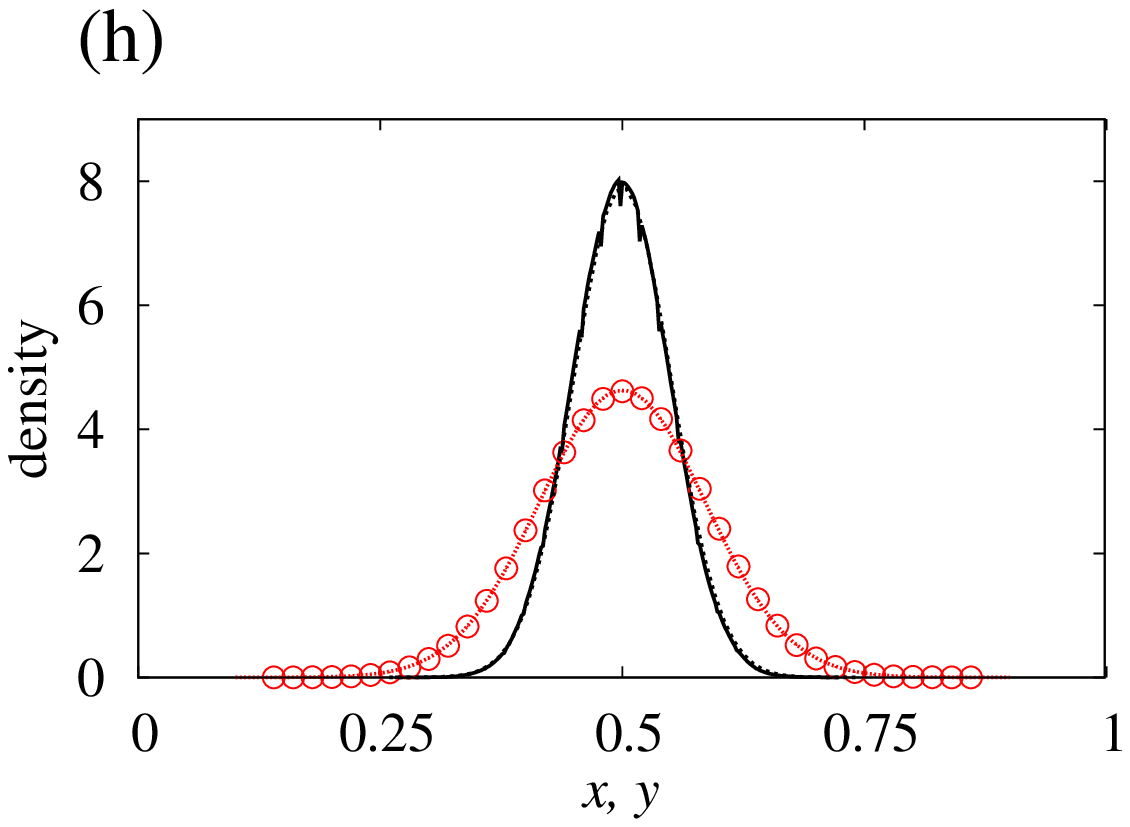}
\includegraphics[width=5cm]{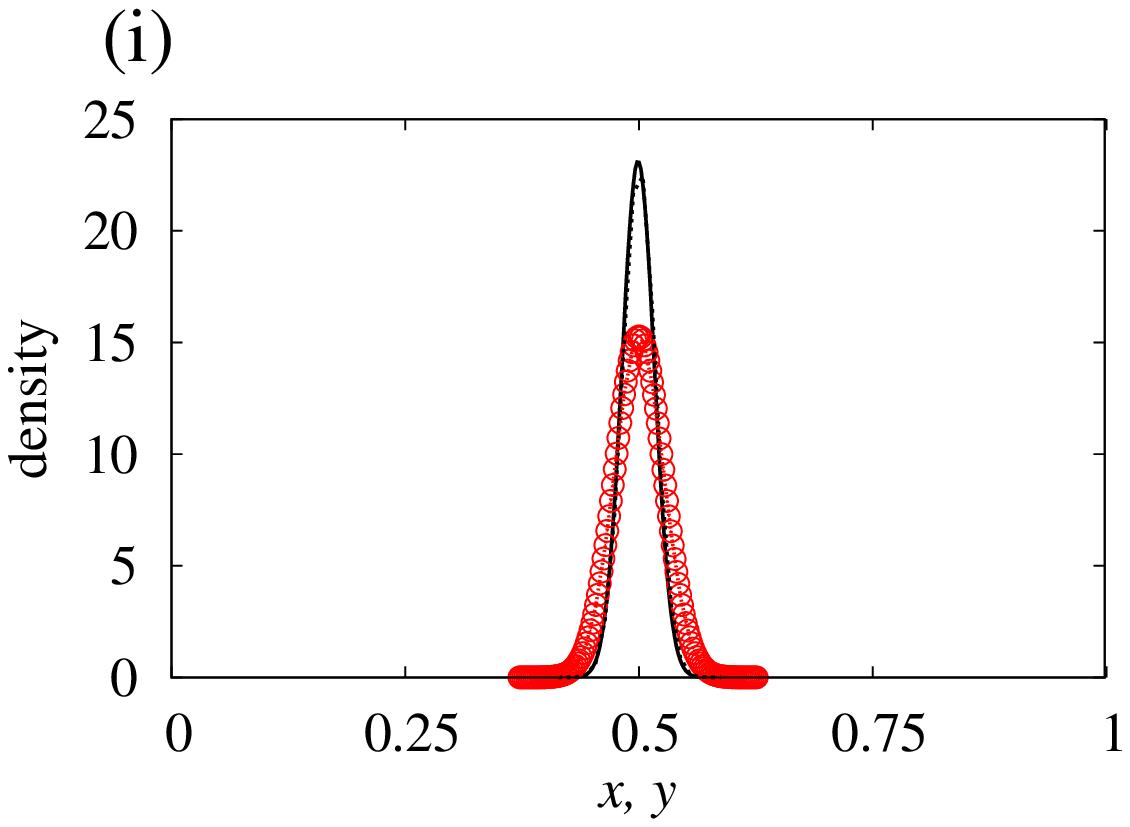}
\caption{The distribution of the fraction of state ${\mathbf 1}$ congregators (i.e., $x$) and that of state ${\mathbf 1}$ contrarians (i.e., $y$) in the equilibrium. I set $N=10000$. (a) Model 1 with $N_y=5$, (b) model 1 with $N_y=50$, (c) model 1 with $N_y=500$, (d) model 2 with $N_y=5$, (e) model 2 with $N_y=50$, (f) model 2 with $N_y=500$, (g) model 3 with $N_y=5$, (h) model 3 with $N_y=50$, and (i) model 3 with $N_y=500$.
In (d) and (e), the distributions are calculated using the results obtained from
$t=0.5\times 10^8$ through $t=10^8$ in a single run,
10 times longer simulation time than in the other cases. This was done
because the convergence of the distributions is much slower in the cases shown in (d) and (e) than in the other cases.}
\label{fig:distribution}
\end{center}
\end{figure}

The results obtained from direct numerical simulations of model 1
are compared with 
the theoretical results given by Eqs.~\eqref{eq:std x1} and \eqref{eq:std y1}
in Fig.~\ref{fig:fluctuation}. I set $N=10000$.
The numerical results agree well with the theory except when $N_y$ is small.
The van Kampen expansion assumes that the relevant distributions are Gaussian.
Numerically calculated distributions of the fraction of congregators in state ${\mathbf 1}$ and that of contrarians in state ${\mathbf 1}$ are compared with the Gaussian distributions with mean 0 and standard deviations as given by
Eqs.~\eqref{eq:std x1} and \eqref{eq:std y1} in Fig.~\ref{fig:distribution}. I set $N=10000$ and examined the cases $N_y=5$ (Fig.~\ref{fig:distribution}(a)),
$N_y=50$ (Fig.~\ref{fig:distribution}(b)), and
$N_y=500$ (Fig.~\ref{fig:distribution}(c)).
The numerically obtained distributions are very close to the theoretical ones
when $N_y$ is not small; in Figs.~\ref{fig:distribution}(b) and \ref{fig:distribution}(c),
the numerical and theoretical results almost completely 
overlap each other for both $x$ and $y$.
In contrast, the numerical and theoretical distributions
are not similar when $N_y$ is small (Fig.~\ref{fig:distribution}(a)).
Discrepancies between the numerical and theoretical results for small $N_y$ values are also nonnegligible in Fig.~\ref{fig:fluctuation}(a).
The deviation
in the case of small $N_y$ owes at least partly to the fact that
the distributions are significantly
affected by the boundary conditions at $x, y=$ 0 and 1. The deviation may be also due to the fact that the distribution of $y$ is very discrete when $N_y$ is small. 

Equations~\eqref{eq:std x1} and \eqref{eq:std y1} imply the following.
First, if the fluctuation of the fraction, not the number, of state ${\mathbf 1}$ congregators and that of state ${\mathbf 1}$ contrarians are compared, they are of the same order. However when the contrarians are rare, $\sigma(x)$ and
$\sigma(y)$ are different by a factor of 3.
 Second, substitution of $N_x=N(1-Y)$ and $N_y=NY$, where $Y$ ($0\le Y\le 1$) is the density of contrarians (section~\ref{sub:rate equations}), in Eqs.~\eqref{eq:std x1} and \eqref{eq:std y1} yields
\begin{align}
\sigma(x) = \sqrt{\frac{1+2Y}{8NY(1-Y)}},\\
\sigma(y) = \sqrt{\frac{3-2Y}{8NY(1-Y)}}.
\end{align}
When $Y$ is fixed, $\sigma(x), \sigma(y) \propto
1/\sqrt{N}$. When $N$ is fixed, it holds that
$\sigma(x), \sigma(y) \propto Y^{-1/2}$ as $Y\to 0$.

\subsubsection{Model 2}

For model 2, the calculations in Appendix~A yield
\begin{align}
\sigma(x) =& \sqrt{\frac{N(N_x+2N_y)}{8N_xN_y^2}}
= \sqrt{\frac{1+Y}{8NY^2(1-Y)}}
\label{eq:std x1 model 2},\\
\sigma(y) =& \sqrt{\frac{N}{8N_y^2}}
= \sqrt{\frac{1}{8NY^2}}.
\label{eq:std y1 model 2}
\end{align}
When $Y$ is fixed, $\sigma(x), \sigma(y) \propto
1/\sqrt{N}$. This result is the same as that for model 1. When $N$ is fixed, it holds that $\sigma(x), \sigma(y) \propto Y^{-1}$
 as $Y\to 0$. This scaling is different from that for model 1. Model 2 generates larger fluctuations than model 1 when the contrarians are rare.

The numerically obtained $\sigma(x)$ and $\sigma(y)$ values are
compared with Eqs.~\eqref{eq:std x1 model 2} and \eqref{eq:std y1
  model 2} in Fig.~\ref{fig:fluctuation}(b). The numerical and
theoretical results agree well when $N_y\ge 100$. It should be noted
that the fluctuation is larger for model 2 than for model 1 when $N_y$
takes intermediate values, i.e., $10\le N_y\le 2000$. The numerically
obtained distributions of $x$ and $y$ are compared with the Gaussian
distributions whose standard deviations are given by
Eqs.~\eqref{eq:std x1 model 2} and \eqref{eq:std y1 model 2} in
Figs.~\ref{fig:distribution}(d), \ref{fig:distribution}(e), and
\ref{fig:distribution}(f) for three $N_y$ values.  The numerical and
theoretical distributions agree well when $N_y$ is large enough (i.e.,
$N_y=500$; Fig.~\ref{fig:distribution}(f)). In Fig.~\ref{fig:distribution}(f), the numerical and
theoretical results almost completely overlap each other for both $x$
and $y$. However, when $N_y$ is smaller, the numerically obtained
distributions of $x$ and $y$ have peaks at $x,y\approx 0$ and 1 such
that they are far from the Gaussian distributions shown by the dotted
lines in Figs.~\ref{fig:distribution}(d) and \ref{fig:distribution}(e). It should be noted that the theoretical results for $x$ and that for $y$ are indistinguishable in Figs.~\ref{fig:distribution}(d) and \ref{fig:distribution}(e).
In this range of $N_y$, the small-fluctuation expansion breaks
down, which is consistent with Fig.~\ref{fig:fluctuation}(b).

\subsubsection{Model 3}

For model 3, the calculations in Appendix~B yield
\begin{align}
\sigma(x) =& \sqrt{\frac{N(N_x+6N_y)}{8N_xN_y(N_x+3N_y)}}
= \sqrt{\frac{1+5Y}{8NY(1-Y)(1+2Y)}}
\label{eq:std x1 model 3},\\
\sigma(y) =& \sqrt{\frac{3N}{8N_y(N_x+3N_y)}}
= \sqrt{\frac{3}{8NY(1+2Y)}}.
\label{eq:std y1 model 3}
\end{align}
When $Y$ is fixed, $\sigma(x), \sigma(y) \propto
1/\sqrt{N}$. When $N$ is fixed, it holds that
$\sigma(x), \sigma(y) \propto Y^{-1/2}$
as $Y\to 0$. The scaling is the same as those for model 1.

The numerically obtained $\sigma(x)$ and $\sigma(y)$ values
are compared with Eqs.~\eqref{eq:std x1 model 3} and \eqref{eq:std y1 model 3} in 
Fig.~\ref{fig:fluctuation}(c). The numerical results agree well with the theory unless $N_y$ is small.
The numerically obtained distributions
of $x$ and $y$ are compared with the Gaussian distributions whose 
standard deviations are given by
Eqs.~\eqref{eq:std x1 model 3} and \eqref{eq:std y1 model 3}
in Figs.~\ref{fig:distribution}(g), \ref{fig:distribution}(h), and
\ref{fig:distribution}(i) for three $N_y$ values.
The theoretical results agree well with the numerical results if $N_y$ is not small
(Figs.~\ref{fig:distribution}(h) and \ref{fig:distribution}(i)).
In Figs.~\ref{fig:distribution}(h) and \ref{fig:distribution}(i), the numerical and theoretical results almost entirely overlap each other.
The results for model 3 are similar to those for model 1.

\subsection{Case of a single contrarian}\label{sub:1 contrarian}

The small-fluctuation approximation cannot capture the behavior of the model when $N_y$ is small (\SEC\ref{sub:van Kampen}). To better understand this situation, I calculate the stationary distribution
of the Fokker-Planck equation for the single-contrarian case, i.e., $N_y=1$. In this extreme case, the single contrarian does not find other contrarians in the population. Therefore, model 1 and model 3 are equivalent. I analyze this model in the following. Model 2 is reduced to the standard voter model and therefore is irrelevant.

There are $N_x=N-1$ congregators. I denote by $P(n_x,0)$ ($P(n_x,1)$) the probability that there are $n_x$ congregators in state ${\mathbf 1}$ and the $n_y=0$ ($n_y=1$) contrarian in 
state ${\mathbf 1}$. The normalization is given by
$\sum_{n_x=0}^{N-1}\left[P(n_x,0)+P(n_x,1) \right]=1$.
The master equations are given by
\begin{align}
N\frac{{\rm d}P(n_x,0)}{{\rm d}t} =& P(n_x,1)\frac{1}{N}\frac{n_x}{N-1}
+ P(n_x-1,0)\frac{(N-1)-(n_x-1)}{N}\frac{n_x-1}{N-1}\notag\\
+& P(n_x+1,0)\frac{n_x+1}{N}\frac{(N-1)-(n_x-1)+1}{N-1}\notag\\
-& P(n_x,0)
\left[
\frac{1}{N}\frac{(N-1)-n_x}{N-1}+\frac{(N-1)-n_x}{N}\frac{n_x}{N-1}+\frac{n_x}{N}\frac{(N-1)-n_x+1}{N-1}
\right],
\label{eq:master 1 single contrarian case}\\
N\frac{{\rm d}P(n_x,1)}{{\rm d}t} =& P(n_x,0)\frac{1}{N}\frac{N-1-n_x}{N-1}
+P(n_x-1,1)\frac{(N-1)-(n_x-1)}{N}\frac{(n_x-1)+1}{N-1}\notag\\
+& P(n_x+1,1)\frac{n_x+1}{N}\frac{(N-1)-(n_x+1)}{N-1}\notag\\
-& P(n_x,1)\left[
\frac{1}{N}\frac{n_x}{N-1}+\frac{(N-1)-n_x}{N}\frac{n_x+1}{N-1}+\frac{n_x}{N}\frac{(N-1)-n_x}{N-1}
\right].
\label{eq:master 2 single contrarian case}
\end{align}
The Fokker-Planck equations on the basis of Eqs.~\eqref{eq:master 1 single contrarian case} and \eqref{eq:master 2 single contrarian case}
are given by
\begin{align}
N\frac{\partial P(n_x,0)}{\partial t} =&
\frac{n_x}{N(N-1)} P(n_x,1)
+ \frac{\partial}{\partial n_x}\left[\frac{n_x}{N(N-1)}
P(n_x,0) \right]\notag\\
&+ \frac{1}{2}\frac{\partial^2}{\partial n_x^2}\left[
\frac{n_x(2N-2n_x-1)}{N(N-1)} P(n_x,0)\right]
-\frac{N-1-n_x}{N(N-1)} P(n_x,0),
\label{eq:FP number 0}\\
N\frac{\partial P(n_x,1)}{\partial t} =&
\frac{N-1-n_x}{N(N-1)} P(n_x,0)
- \frac{\partial}{\partial n_x}\left[
\frac{N-1-n_x}{N(N-1)} P(n_x,1)\right]\notag\\
&+\frac{1}{2}\frac{\partial^2}{\partial n_x^2}
\left[ \frac{(N-1-n_x)(2n_x+1))}{N(N-1)} P(n_x,1)
\right]
- \frac{n_x}{N(N-1)} P(n_x,1).
\label{eq:FP number 1}
\end{align}
In terms of the fraction of state ${\mathbf 1}$ congregators in the congregator subpopulation, i.e., $x$, Eqs.~\eqref{eq:FP number 0} and
\eqref{eq:FP number 1} are given by
\begin{align}
N^2\frac{\partial P(x,0)}{\partial t} =&
x P(x,1)-(1-x) P(x,0)\notag\\ 
&+ \frac{1}{N-1}\frac{\partial}{\partial x}\left[x P(x,0)\right]
+\frac{1}{N-1}\frac{\partial^2}{\partial x^2}\left\{x\left[1-x+\frac{1}{2(N-1)}\right] P(x,0)\right\},
\label{eq:FP fraction 0}\\
N^2\frac{\partial P(x,1)}{\partial t} =&
(1-x) P(x,0)-x P(x,1)\notag\\
&- \frac{1}{N-1}\frac{\partial}{\partial x}\left[(1-x) P(x,1)\right]
+\frac{1}{N-1}\frac{\partial^2}{\partial x^2}\left\{\left[x+\frac{1}{2(N-1)}\right](1-x) P(x,1)\right\}.
\label{eq:FP fraction 1}
\end{align}

When $N$ is large, $P(x,0)$ and $P(x,1)$
evolve on a fast timescale until
\begin{equation}
x P(x,1)-(1-x) P(x,0) = \frac{g(x)}{N-1}
\label{eq:fast timescale}
\end{equation}
is satisfied, where $g(x)=O(1)$. Although I am interested in the equilibrium, the adiabatic approximation given by Eq.~\eqref{eq:fast timescale} holds true in the course of the dynamics on a slow timescale as well as in the equilibrium. By substituting
Eq.~\eqref{eq:fast timescale} in 
Eqs.~\eqref{eq:FP fraction 0} and \eqref{eq:FP fraction 1}, I obtain the following equations in the equilibrium:
\begin{align}
g(x) + \frac{\partial}{\partial x}\left[\frac{x^2}{1-x} P(x,1)
-\frac{x}{(N-1)(1-x)}g(x) \right]\notag\\
+ \frac{\partial^2}{\partial x^2}
\left\{\left[1+\frac{1}{2(1-x)(N-1)}\right]
\left[x^2 P(x,1)-\frac{x}{N-1}g(x) \right] \right\} =& 0,\\
-g(x)-\frac{\partial}{\partial x}\left[(1-x) P(x,1)\right] 
+ \frac{\partial^2}{\partial x^2}
\left\{\left[x+\frac{1}{2(N-1)}\right](1-x) P(x,1) \right\} =& 0.
\end{align}
By ignoring $O[1/(N-1)]$ terms, which is justified unless $x=O(1/N)$ or 
$1-x=O(1/N)$,
I obtain
\begin{align}
g(x) + \frac{\partial}{\partial x}\left[\frac{x^2}{1-x} P(x,1)\right]
+ \frac{\partial^2}{\partial x^2}
\left[x^2 P(x,1) \right]=& 0,
\label{eq:approximated equilibrium 1}\\
-g(x)-\frac{\partial}{\partial x}\left[(1-x) P(x,1)\right] 
+ \frac{\partial^2}{\partial x^2}
\left[x(1-x) P(x,1) \right] =& 0.
\label{eq:approximated equilibrium 2}
\end{align}
By summing Eqs.~\eqref{eq:approximated equilibrium 1}
and \eqref{eq:approximated equilibrium 2}, I obtain
\begin{equation}
\frac{\partial}{\partial x}\left[\frac{-1+2x}{1-x} P(x,1) \right]
+ \frac{\partial^2}{\partial x^2}\left[x P(x,1)\right]=0.
\label{eq:summed approximated equilibrium}
\end{equation}
The general solution of Eq.~\eqref{eq:summed approximated equilibrium} is given by
\begin{equation}
P(x,1)= C_1(1-x)\log\frac{x}{1-x} + C_2(1-x),
\label{eq:general solution}
\end{equation}
where $C_1$ and $C_2$ are constants. Equation~\eqref{eq:general solution} and
the symmetry relationship
$P(x,0)=P(1-x,1)$ yield
\begin{align}
x P(x,1)-(1-x) P(x,0) =& x P(x,1)-(1-x) P(1-x,1)\notag\\
=& 2C_1x(1-x)\log\frac{x}{1-x}.
\end{align}
For this quantity to be of order $O(1/N)$ (see Eq.~\eqref{eq:fast timescale}), $C_1=0$ is required. It should be noted that I have already discarded $O(1/N)$ terms in deriving Eqs.~\eqref{eq:approximated equilibrium 1}
and \eqref{eq:approximated equilibrium 2}.
Therefore, Eq.~\eqref{eq:general solution} is reduced to $P(x,1)=C_2(1-x)$. The normalization condition $\int_0^1 P(x,1)dx=1/2$, which in fact should be applied with a caution because the solution given by
Eq.~\eqref{eq:general solution}
may be invalid near $x=0$ and $x=1$, leads to $C_2=1$. Finally, I obtain
\begin{align}
P(x,0)=& x,\label{eq:solution 1}\\
P(x,1)=& 1-x.\label{eq:solution 2}
\end{align}

\begin{figure}
\begin{center}
\includegraphics[width=8cm]{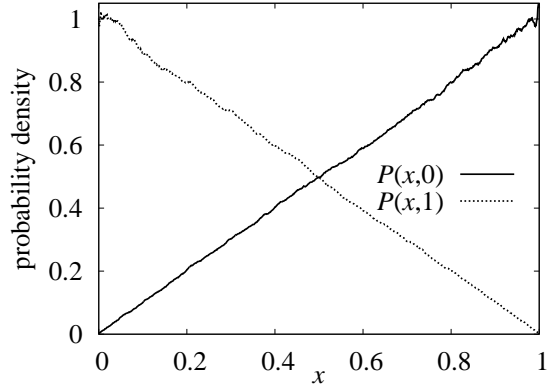}
\caption{Distribution of the fraction of ${\mathbf 1}$ congregators when $N_y=1$. The distribution conditioned that the contrarian is in state ${\mathbf 0}$
and that conditioned that the contrarian is in state ${\mathbf 1}$
are shown by the solid and dotted lines, respectively. $N=1000$.}
\label{fig:single contrarian}
\end{center}
\end{figure}

Equations~\eqref{eq:solution 1} and \eqref{eq:solution 2} imply that the fraction of ${\mathbf 1}$ congregators that is not conditioned by the state of the contrarian, i.e., $P(x,0)+P(x,1)$,
 is uniformly distributed on $[0,1]$. These equations also imply that
the congregators would be in the ${\mathbf 0}$ state
when the contrarian is in the ${\mathbf 1}$ state. This phenomenon occurs because the contrarian tries to escape from the congregators. Numerically obtained equilibrium distributions are shown in Fig.~\ref{fig:single contrarian} for $N=1000$. The results are in an excellent agreement with Eqs.~\eqref{eq:solution 1} and \eqref{eq:solution 2}.

The uniform distribution of $x$ implies $\sigma(x)=1/(2\sqrt{3})$. This value is in fact approached as $N_y$ is decreased in models 1 and 3 (squares in \FIGS\ref{fig:fluctuation}(a) and \ref{fig:fluctuation}(c)).

\section{Discussion}

I proposed extensions of the voter model with contrarian agents. Even a single contrarian turns out to change the final configuration of the dynamics from the consensus of one state to the coexistence of the two states. Among the three models analyzed in the present study, model 2 behaves differently from models 1 and 3 in that the coexistence equilibrium is much less stable in model 2 than models 1 and 3 when contrarians are rare. This difference is likely to
owe to the fact that contrarians are assumed to like congregators only in model 2 (Table~\ref{tab:3 models def}). The results that model 1 and model 3 behave similarly suggest that the behavior of contrarians toward the conspecific does not much affect the collective behavior of the model, at least in the current framework. It should be noted that, with model 2 included, the collective dynamics of the model is robust with respect to the behavioral rule of the agent (Table~\ref{tab:3 models def}) if there are sufficient contrarians in the population.

The effect of contrarians has been investigated in various models of opinion formation, as reviewed in \SEC\ref{sec:introduction}. Most of the previous models also found that contrarians promote coexistence of different states when consensus is inevitable in the models without contrarians. The strength of the current study lies in that I confined myself to linear models, as is the original voter models, and reached strong analytical conclusions. I used the van Kampen small-fluctuation approximations and solved the case of a single contrarian to characterize the fluctuations around the coexistence equilibrium. As the number of contrarians increases, the equilibrium distributions change from the uniform distribution to the Gaussian distribution with small standard deviations. It should be also noted that the present models do not show phase transition, whereas nonlinear models usually show phase transitions between the consensus-like phase
and the coexistence phase.

The voter model is peculiar in the sense that it is linear and thus without assumed threshold behavior (see, for example, Refs.~\cite{Galam1998EPJB,Galam2005EPL,Galam2008IJMPC,Castellano2009RMP} for nonlinear models).
In contrast to the present models, nonlinear opinion formation models with contrarians can show
rich behavior. For example,
periodic and chaotic behavior is briefly described
in a model with contrarians
constructed on the basis of the Ising spin system \cite{Kurten2008IJMPB}.
In the context of nonlinear coupled phase oscillators, rich behavior including traveling waves and partial synchrony was reported
\cite{HongStrogatz2011PRL}. Further investigating
nonlinear as well as linear opinion formation models with contrarians
warrants future work.

\section*{Appendix A: Small-fluctuation approximation for model 2}

For model 2, the master equation in terms of $P$, the increment
operators, and decrement operators is given by
\begin{align}
N\frac{{\rm d}P}{{\rm d}t} =&
(E_x-1)
\left[n_x\left(N_x-n_x+N_y-n_y\right)P\right]
+(E_x^{-1}-1)\left[\left(N_x-n_x\right)\left(n_x+n_y\right)P\right]\notag\\
+&(E_y-1)\left[n_y\left(N_x-n_x+n_y\right)P\right]
+(E_y^{-1}-1)\left[\left(N_y-n_y\right)\left(n_x+N_y-n_y\right)P\right].
\label{eq:master equation with P model 2}
\end{align}
Substitution of Eqs.~\eqref{eq:n_x ansatz}, \eqref{eq:n_y ansatz},
\eqref{eq:E_x}, \eqref{eq:E_x^-}, \eqref{eq:E_y}, and \eqref{eq:E_y^-}
in Eq.~\eqref{eq:master equation with P model 2} yields
\begin{align}
& N\left(\frac{\partial\Pi}{\partial t} - \sqrt{N_x}\frac{{\rm d}x}{{\rm d}t}
\frac{\partial\Pi}{\partial\xi} - \sqrt{N_y}\frac{{\rm d}y}{{\rm d}t}
\frac{\partial\Pi}{\partial\eta} \right)\notag\\
=& \left(\frac{1}{\sqrt{N_x}}\frac{\partial}{\partial\xi}
+\frac{1}{2N_x}\frac{\partial^2}{\partial\xi^2} \right)
\left(N_x x+\sqrt{N_x}\xi\right)
\left[N_x\left(1-x\right)-\sqrt{N_x}\xi
+N_y\left(1-y\right)-\sqrt{N_y}\eta\right]\Pi\notag\\
&+
 \left(-\frac{1}{\sqrt{N_x}}\frac{\partial}{\partial\xi}
+\frac{1}{2N_x}\frac{\partial^2}{\partial\xi^2} \right)
\left[N_x\left(1-x\right)-\sqrt{N_x}\xi\right]
\left(N_x x+\sqrt{N_x}\xi
+N_y y+\sqrt{N_y}\eta\right)\Pi\notag\\
&+ \left(\frac{1}{\sqrt{N_y}}\frac{\partial}{\partial\eta}
+\frac{1}{2N_y}\frac{\partial^2}{\partial\eta^2} \right)
\left(N_yy+\sqrt{N_y}\eta\right)
\left[N_x\left(1-x\right)-\sqrt{N_x}\xi
+N_y y+\sqrt{N_y}\eta\right]\Pi\notag\\
&+ \left(-\frac{1}{\sqrt{N_y}}\frac{\partial}{\partial\eta}
+\frac{1}{2N_y}\frac{\partial^2}{\partial\eta^2} \right)
\left[N_y\left(1-y\right)-\sqrt{N_y}\eta\right]
\left(N_x x+\sqrt{N_x}\xi
+N_y\left(1-y\right)-\sqrt{N_y}\eta\right)\Pi. 
\label{eq:master equation with Pi model 2}
\end{align}

The highest order terms of Eq.~\eqref{eq:master equation with Pi model
  2} recovers the mean-field dynamics given by Eqs.~\eqref{eq:x} and \eqref{eq:y model 2}.
The comparison of the second highest order terms in 
Eq.~\eqref{eq:master equation with Pi model 2} yields
\begin{align}
N\frac{\partial\Pi}{\partial t} =&
N_y\frac{\partial}{\partial\xi}(\xi\Pi)
-\sqrt{N_xN_y}\eta\frac{\partial\Pi}{\partial\xi}
+
\left[N_x x(1-x)+\frac{N_y}{2}(x+y-2xy)\right]
\frac{\partial^2\Pi}{\partial\xi^2}\notag\\
-&\sqrt{N_xN_y}\xi\frac{\partial\Pi}{\partial\eta}
+\left(N_x+2N_y\right)\frac{\partial}{\partial\eta}(\eta\Pi)
+\left[\frac{N_x}{2}(x+y-2xy)+ \frac{N_y}{2}(1-2y+2y^2)\right]\frac{\partial^2\Pi}{\partial\eta^2}.
\label{eq:2nd highest model 2}
\end{align}
Application of $\int\int{\rm d}\xi {\rm d}\eta\,\xi$ and
$\int\int{\rm d}\xi {\rm d}\eta\,\eta$ to
Eq.~\eqref{eq:2nd highest model 2} yields
Eq.~\eqref{eq:d<xi>/dt model 1} and
\begin{equation}
N\frac{\partial\left<\eta\right>}{\partial t} = 
\sqrt{N_xN_y}\left<\xi\right> - \left(N_x+2N_y\right)\left<\eta\right>,
\label{eq:d<eta>/dt model 2}
\end{equation}
respectively.
Because the characteristic equation of the 
Jacobian of the dynamics given by Eqs.~\eqref{eq:d<xi>/dt
  model 1} and \eqref{eq:d<eta>/dt model 2} coincides with
Eq.~\eqref{eq:eigen eq model 2},
the dynamics converges to the unique equilibrium given by
$\left<\xi\right>^* = \left<\eta\right>^*=0$.

Application of $\int\int{\rm d}\xi {\rm d}\eta\,\xi^2$,
$\int\int{\rm d}\xi {\rm d}\eta\,\xi\eta$, and
$\int\int{\rm d}\xi {\rm d}\eta\,\eta^2$ to
 Eq.~\eqref{eq:2nd highest model 2} yields
Eq.~\eqref{eq:d<xi^2>/dt model 1},
\begin{align}
N\frac{\partial\left<\xi\eta\right>}{\partial t} =&
\sqrt{N_xN_y}\left<\xi^2\right>
-(N_x+3N_y)\left<\xi\eta\right>+\sqrt{N_xN_y}\left<\eta^2\right>,
\label{eq:d<xi*eta>/dt model 2}
\end{align}
and
\begin{align}
N\frac{\partial\left<\eta^2\right>}{\partial t} =&
N_x(x+y-2xy) + N_y(1-2y+2y^2)
 +2\sqrt{N_xN_y}\left<\xi\eta\right>
-2(N_x+2N_y)\left<\eta^2\right>,
\label{eq:d<eta^2>/dt model 2}
\end{align}
respectively. By substituting
$(x^*, y^*) = (1/2, 1/2)$ in
Eqs.~\eqref{eq:d<xi^2>/dt model 1}, \eqref{eq:d<xi*eta>/dt model 2},
and \eqref{eq:d<eta^2>/dt model 2}
and setting the left-hand sides to 0, I obtain
\begin{align}
\left<\xi^2\right>^* =& \frac{N(N_x+2N_y)}{8N_y^2},
\label{eq:<xi^2>^* model 2}\\
\left<\xi\eta\right>^* =& \frac{N\sqrt{N_x}}{8N_y^{\frac{3}{2}}},
\label{eq:<xi*eta>^* model 2}
\end{align}
and
\begin{align}
\left<\eta^2\right>^* =& \frac{N}{8N_y}.
\label{eq:<eta^2>^* model 2}
\end{align}
By using the relationship 
$\sigma(x) =\sqrt{\left<\xi^2\right>^*/N_x}$ and
$\sigma(y) =\sqrt{\left<\eta^2\right>^*/N_y}$,
I obtain Eqs.~\eqref{eq:std x1 model 2} and \eqref{eq:std y1 model 2}.

\section*{Appendix B: Small-fluctuation approximation for model 3}

For model 3, the master equation is given by
\begin{align}
N\frac{{\rm d}P}{{\rm d}t} =&
(E_x-1)
\left[n_x\left(N_x-n_x+N_y-n_y\right)P\right]
+(E_x^{-1}-1)\left[\left(N_x-n_x\right)\left(n_x+n_y\right)P\right]\notag\\
+&(E_y-1)\left[n_y\left(n_x+n_y\right)P\right]
+(E_y^{-1}-1)\left[\left(N_y-n_y\right)\left(N_x-n_x+N_y-n_y\right)P\right].
\label{eq:master equation with P model 3}
\end{align}
Substitution of Eqs.~\eqref{eq:n_x ansatz}, \eqref{eq:n_y ansatz},
\eqref{eq:E_x}, \eqref{eq:E_x^-}, \eqref{eq:E_y}, and \eqref{eq:E_y^-}
in Eq.~\eqref{eq:master equation with P model 3} yields
\begin{align}
& N\left(\frac{\partial\Pi}{\partial t} - \sqrt{N_x}\frac{{\rm d}x}{{\rm d}t}
\frac{\partial\Pi}{\partial\xi} - \sqrt{N_y}\frac{{\rm d}y}{{\rm d}t}
\frac{\partial\Pi}{\partial\eta} \right)\notag\\
=& \left(\frac{1}{\sqrt{N_x}}\frac{\partial}{\partial\xi}
+\frac{1}{2N_x}\frac{\partial^2}{\partial\xi^2} \right)
\left(N_x x+\sqrt{N_x}\xi\right)
\left[N_x\left(1-x\right)-\sqrt{N_x}\xi
+N_y\left(1-y\right)-\sqrt{N_y}\eta\right]\Pi\notag\\
&+
 \left(-\frac{1}{\sqrt{N_x}}\frac{\partial}{\partial\xi}
+\frac{1}{2N_x}\frac{\partial^2}{\partial\xi^2} \right)
\left[N_x\left(1-x\right)-\sqrt{N_x}\xi\right]
\left(N_xx+\sqrt{N_x}\xi
+N_yy+\sqrt{N_y}\eta\right)\Pi\notag\\
&+ \left(\frac{1}{\sqrt{N_y}}\frac{\partial}{\partial\eta}
+\frac{1}{2N_y}\frac{\partial^2}{\partial\eta^2} \right)
\left(N_y y+\sqrt{N_y}\eta\right)
\left(N_x x+\sqrt{N_x}\xi
+N_y y+\sqrt{N_y}\eta\right)\Pi\notag\\
&+ \left(-\frac{1}{\sqrt{N_y}}\frac{\partial}{\partial\eta}
+\frac{1}{2N_y}\frac{\partial^2}{\partial\eta^2} \right)
\left[N_y\left(1-y\right)-\sqrt{N_y}\eta\right]
\left[N_x\left(1-x\right)-\sqrt{N_x}\xi
+N_y\left(1-y\right)-\sqrt{N_y}\eta\right]\Pi. 
\label{eq:master equation with Pi model 3}
\end{align}

The highest order terms of Eq.~\eqref{eq:master equation with Pi model 3} recovers the mean-field dynamics given by Eqs.~\eqref{eq:x} and \eqref{eq:y model 3}.
The comparison of the second highest order terms in 
Eq.~\eqref{eq:master equation with Pi model 3} yields
\begin{align}
N\frac{\partial\Pi}{\partial t} =&
N_y\frac{\partial}{\partial\xi}(\xi\Pi)
-\sqrt{N_xN_y}\eta\frac{\partial\Pi}{\partial\xi}
+
\left[N_x x(1-x)+\frac{N_y}{2}(x+y-2xy)\right]
\frac{\partial^2\Pi}{\partial\xi^2}\notag\\
+&\sqrt{N_xN_y}\xi\frac{\partial\Pi}{\partial\eta}
+(N_x+2N_y)\frac{\partial}{\partial\eta}(\eta\Pi)
+\left[\frac{N_x}{2}(1-x-y+2xy)+ \frac{N_y}{2}(1-2y+2y^2)\right]\frac{\partial^2\Pi}{\partial\eta^2}.
\label{eq:2nd highest model 3}
\end{align}
Application of $\int\int{\rm d}\xi {\rm d}\eta\,\xi$ and
$\int\int{\rm d}\xi {\rm d}\eta\,\eta$ to
Eq.~\eqref{eq:2nd highest model 3} yields
Eq.~\eqref{eq:d<xi>/dt model 1} and
\begin{equation}
N\frac{\partial\left<\eta\right>}{\partial t} = 
- \sqrt{N_xN_y}\left<\xi\right> - (N_x+2N_y)\left<\eta\right>,
\label{eq:d<eta>/dt model 3}
\end{equation}
respectively.
Because the characteristic equation of the 
Jacobian of the dynamics given by Eqs.~\eqref{eq:d<xi>/dt
  model 1} and \eqref{eq:d<eta>/dt model 3} coincides with
Eq.~\eqref{eq:eigen eq model 3},
the dynamics converges to $\left<\xi\right>^* = \left<\eta\right>^*=0$.

Application of $\int\int{\rm d}\xi {\rm d}\eta\,\xi^2$,
$\int\int{\rm d}\xi {\rm d}\eta\,\xi\eta$, and
$\int\int{\rm d}\xi {\rm d}\eta\,\eta^2$
to Eq.~\eqref{eq:2nd highest model 3} yields
Eq.~\eqref{eq:d<xi^2>/dt model 1},
\begin{align}
N\frac{\partial\left<\xi\eta\right>}{\partial t} =&
-\sqrt{N_xN_y}\left<\xi^2\right>
-(N_x+3N_y)\left<\xi\eta\right>+\sqrt{N_xN_y}\left<\eta^2\right>,
\label{eq:d<xi*eta>/dt model 3}
\end{align}
and
\begin{align}
N\frac{\partial\left<\eta^2\right>}{\partial t} =&
N_x(1-x-y+2xy) + N_y(1-2y+2y^2)
-2\sqrt{N_xN_y}\left<\xi\eta\right>
-2(N_x+2N_y)\left<\eta^2\right>,
\label{eq:d<eta^2>/dt model 3}
\end{align}
respectively. By substituting
$(x^*, y^*) = (1/2, 1/2)$ in
Eqs.~\eqref{eq:d<xi^2>/dt model 1}, \eqref{eq:d<xi*eta>/dt model 3},
and \eqref{eq:d<eta^2>/dt model 3}
and setting the left-hand sides to 0, I obtain
\begin{align}
\left<\xi^2\right>^* =& \frac{N(N_x+6N_y)}{8N_y(N_x+3N_y)},
\label{eq:<xi^2>^* model 3}\\
\left<\xi\eta\right>^* =& -\frac{N\sqrt{N_x}}{8(N_x+3N_y)\sqrt{N_y}},
\label{eq:<xi*eta>^* model 3}
\end{align}
and
\begin{align}
\left<\eta^2\right>^* =& \frac{3N}{8(N_x+3N_y)},
\label{eq:<eta^2>^* model 3}
\end{align}
which lead to
Eqs.~\eqref{eq:std x1 model 3} and \eqref{eq:std y1 model 3}.

\begin{acknowledgments}
I thank Sidney Redner for initial collaboration on this project. 
I thank Shoma Tanabe for
careful reading of the paper.
I acknowledge the support provided by
the Grants-in-Aid for Scientific Research
(Grant No. 20540382) from MEXT, Japan,
the Nakajima Foundation, and JSPS and F.R.S.-FNRS under the Japan-Belgium Research Cooperative Program.
\end{acknowledgments}


%

\end{document}